\documentclass{ws-ijmpa}
\usepackage{slashed}
\newcommand{\mtstar}{{\ensuremath{m_{T}^{\star}}}}
\newcommand{\mttrue}{{\ensuremath{m_{T}^{\mathrm{true}}}}}
\newcommand{\mtbound}{{\ensuremath{m_{T}^{\mathrm{bound}}}}}
\begin{document}

\markboth{Ben Gripaios}
{Tools for extracting new physics in events with missing transverse momentum}

%
\catchline{}{}{}{}{}
%

\title{Tools for extracting new physics in events with missing
  transverse momentum
}

\author{BEN GRIPAIOS
}

\address{Cavendish Laboratory, University of Cambridge,\\
J J Thomson Avenue, Cambridge, CB3 0HE,
United Kingdom.\\
gripaios@hep.phy.cam.ac.uk}

\maketitle

\begin{history}
\received{Day Month Year}
\revised{Day Month Year}
\end{history}

\begin{abstract}
We review tools that have been developed in recent years to maximize
our ability to discover and characterize new physics appearing in LHC
events with missing transverse momentum.
\end{abstract}

\section{Introduction, Aims and Scope}	
The 7 TeV run of the Large Hadron Collider (LHC) is now well underway
and a plethora of searches for new physics has already been carried
out.

This short review concerns searches for new physics in LHC events with
missing transverse
momentum.\footnote{Unfortunately, the alternative moniker ``missing
  energy'', which is widely used in the literature, may refer to any
  one of a multitude of different observables;\cite{Barr:2011xt} here we insist on using
``missing transverse momentum'', whose definition is, we hope, unambiguous.}
Many of these searches are based on techniques that have been
developed in the last few years, specifically for the advent of the
LHC, and the main purpose of this review is to describe some of those
techniques and the principles that underlie them. 

The review will also cover developments in the strategies that are
planned to measure the properties of any new physics that we may be
lucky enough to discover. This characterization of new physics is not
wholly divorced from the discovery process itself, but the two
processes are sufficiently different that
they merit separate consideration.

The experimental analyses with which this review concerns itself, like the machine and detectors themselves,
are imponderably complicated. This review contains no discussion
whatsoever of issues related to hardware, calibration, particle
isolation and identification, triggering, jet reconstruction, pile-up,
underlying event, and so
on. Instead, the starting point will be a set of high-level objects
(jets, leptons, missing transverse momentum, etc) in an event, together with their
associated measured momenta and uncertainties thereon. Given these, we
shall ask which observable or observables, considered as functions of the
momenta (visible or missing), optimize our ability to either make a discovery of new
physics or to measure some property thereof.

Even with such a narrow focus, the discussion will, necessarily,
be incomplete. The literature is, moreover, large and ever-growing, so that
one cannot even hope to catalogue it in a definitive fashion. Rather,
the hope is to provide, at least, a useful, if idiosyncratic, introduction to
the subject and its methodology. 

Before closing this introductory section, it is perhaps useful to spell out the reasons
for concentrating on events with missing transverse momentum.
One is that methods for dealing for LHC events without missing transverse momentum can
often be straightforwardly copied from those used previously at hadron
colliders or elsewhere.\footnote{It is certainly true, however, that
the LHC, which operates in a previously uncharted regime of energy and
luminosity, creates challenges and opportunities of its own, even in
events without missing transverse momentum. As a
simple example, the fact that typical partonic collision CM energies
at the LHC greatly exceed the masses of SM particles leads to an
abundance of events with highly-boosted objects in the final state,
whose decay products are well-collimated. This has led to the
development of jet substructure techniques, reviewed in Ref.~\refcite{Abdesselam:2010pt}.} 
To take an example which we shall discuss in more detail below, in
searching for new resonances it suffices to ``blindly'' compute invariant mass distributions of some
combination of final state momenta. But in events with missing transverse momentum, we
shall argue that the optimal observable strongly depends on the details of the
signal dynamics. So, even though we have, by now, a great deal of
experience in studying missing transverse momentum events (for example, in the
discoveries of the $W$-boson and top quark at CERN and Fermilab
respectively) , we still have to work hard to be sure that we have
optimized our ability to discover physics at the LHC whose nature
is not yet known to us.

A second reason, of course, is that many interesting scenarios of new
physics, both within the
Standard Model (SM) and beyond it, do predict missing transverse
momentum events at the LHC. Indeed, we can expect to see missing transverse momentum whenever a neutrino is produced in the decay of a Higgs boson. Moreover, we hope
to be able to produce new, invisible particles at the LHC, most
notably Dark Matter.

A third reason is purely pragmatic. Data is produced in LHC collisions
at such a rate that one must either throw most of it away at random
with a ``pre-scaled'' trigger,
or one must trigger on some characteristic. One suitable characteristic is
missing transverse momentum, not least because its presence tells us that the event
cannot be some tedious QCD process, which overwhelmingly dominate the
total cross-section. Unfortunately, most of the time such events will
be QCD, but badly measured, and indeed a large part of the remaining
work will be to separate this background from a signal of events with
genuine missing transverse momentum.

The outline is as follows. In the next Section, we discuss the issue
of finding an optimal observable and why there is a necessity to do so
in the first place. In Section \ref{sec:kinematics}, we consider one
approach to finding such an optimal
observable\footnote{Needless to say, the notion of which observable is
optimal depends on one's exact criterion for what is optimal.} in the context of what is perhaps the most simple example,
namely leptonic decays of the $W$-boson.  We then generalize the same
methodology, which essentially amounts to the assumption that the most
important distinguishing feature of the signal (compared to
backgrounds) is its kinematics, to more complicated situations, including decays of a
Higgs boson to $W$-bosons or $\tau$-mesons and pair production of
supersymmetric partners of SM particles.

The $h \rightarrow \tau \tau$ example is particularly appropriate for the
purposes of our general discussion, since a number of more sophisticated approaches seem to be
possible. Even focussing only on kinematics, it is possible to fully
reconstruct events at the LHC, up to a discrete ambiguity, meaning
that one can, essentially, carry over
the same bump hunting methods used for events without missing transverse momentum.
But $h \rightarrow \tau \tau$ searches are also interesting because
they give real-world examples of analyses which try to go beyond
kinematics and try to use dynamical information to discriminate signal
from background. We discuss analyses proposed or in use by both ATLAS
and CMS in Section \ref{sec:dynamics}. We then go on to consider how
assumptions about the dynamics may help in SUSY searches, in a variety
of ways.

Finally, we turn to the issue of measuring the properties of new
physics, such as masses, spins, and couplings. There already exist
reviews covering these subjects, so our discussion will be brief. We also discuss more
generally the problem of how we can make inferences about Dark Matter
on the basis of LHC measurements and how best to characterize new
physics in its nascent stages.
\section{Optimal Observables \label{sec:optimal}	}
To begin with, one might wonder why there is any need to devote a
review to discussing optimal observables for events with missing transverse momentum.
Indeed, the question of `Which observable is
optimal?' was answered for us long ago by the statisticians. The
answer is that the optimal observable is given by the likelihood
function. In particle collisions, this amounts to the matrix element
for a particular scattering process, including signal and background
contributions, convoluted with the appropriate detector response and
marginalized with respect to quantities that go unmeasured.

Unfortunately, this observable is
nigh-on impossible to compute in practice: not only is our modelling
of the matrix element (for either signal or background) and detector response insufficient, but also the
computational effort required to carry out such an analysis at the LHC
is inconceivable. 

Even then, one might wonder why there is any need for phenomenologists
to scratch their heads trying to find an optimal observable. At
least when one is dealing with high-level objects after event selection, any observable may
be written as a function of only a few basic observable momenta. 
So one might object that again the statisticians have already done the
hard work
in designing multivariate analyses, such as neural networks,
boosted decision trees, etc, which will find the optimal observable
for us, once we have trained them up sufficiently. 

There are two counter-objections to this objection. The first, as de R\'{u}jula
and Galindo\cite{Rujula:2011qn} so charmingly put it, is that there are times when one
would like to have the ``pleasure of understanding with use of one's own neural network'' in order
to have a clear picture of what is going on. The second is that
the sophisticated multivariate analyses on the market are perhaps not
so sophisticated, in that they typically only form linear combinations
of the initial observables. As such, they miss variables constructed,
for example, as products or ratios of the initial observables. Indeed,
one does not have to look too hard to find examples in the recent literature where humans have
succeeded in finding useful observables where multivariate methods
have failed.\cite{Thaler:2011gf} So, {\em nolens volens}, there is still work for us to do. 

Even though the true likelihood function is unattainable, it does suggest a
useful way to approach the problem of finding an optimal observable.
Indeed,
imagine that one is able to identify which aspect or aspects of the physics
most affects our ability to make a discovery or a measurement in a particular
channel. One could then try to find the observable that best takes
account of this aspect. In the next Section, we shall develop this
idea in the context of several examples in which it is the differing kinematic
properties of the signal and background which are assumed to be paramount.
\section{Kinematics \label{sec:kinematics}}	
We have already mentioned how one might
search for a resonance in events without missing transverse momentum. The tried and trusted way to do so is
to plot the invariant mass of some combination of final state
particles and to look for a bump. This is a useful thing to do, not only because Lorentz-invariance guarantees that signal events
will pile up at the resonance mass (increasing the discovery
potential), but also because
 it is rather hard to imagine effects coming from
smoothly-varying backgrounds
and detector response which could give rise to a bump.\footnote{In
  contrast, it is rather easy to obtain features like ``tails'' and
  ``shoulders'' in distributions by superposing simple background
  components and caution should be exercised in claiming discoveries
  on the basis of such features.} Thus, even with
almost complete ignorance about the nature or size of the backgrounds
and detector response, one might confidently claim a discovery if a bump
were seen.  

In this example, the aspect of the physics which is most important is
the (assumed) presence of a narrow resonance peak in the invariant
mass distribution of the signal and so
the natural choice of observable is the invariant mass. The assumption
is optimal\footnote{We stress that we have not proved that the
  observable is optimal, not least because we have not defined what we
  mean by ``optimal''. We use it in a colloquial sense.} not only for discovery, but also for measuring
the mass of the resonance, since signal events pile up at the place
where the measurement is performed. 

In a sense, this way of proceeding is akin to following the
likelihood
approach, albeit in a cavalier fashion. In effect, one is making the drastic approximation that ``the physics''
is encapsulated by the chosen aspect (kinematics) and nothing else and
then making a crude guess as to the resulting
likelihood variable.\footnote{Presumably, one could follow this recipe
  in a rigorous fashion, but we shall not do so here.}

It is important to note that making drastic approximations of this
kind
cannot invalidate the subsequent analysis,
even if the approximations made are badly
violated in reality. Indeed, the only purpose of the procedure is to
define (or rather, motivate) an
observable. But some observables are better than others and so the price that one
pays for a poor approximation is an observable that is far from being
optimal.  

So, one can already can make a lot of progress just by using kinematic features of
the signal, choosing an observable whose signal distribution is as
``peaked'' (or piled-up) as possible. Of course, in order for this to
be a useful course of action, one should
take care that the background not be equally sculpted at the same time!

This idea, as we will see, generalizes in a very natural way to
events with missing transverse momentum. When one (or more) of the final state
particles is (or are) invisible, we cannot hope to construct a resonance
bump in an invariant mass, since the invariant mass is no longer
observable. But we can try to construct an observable for which the
signal is as peaked as possible. To do so, let us begin with the case where all final state
particles are visible,  for which the invariant mass observable is always equal to
the resonance mass. To put it another way, the observable is bounded both above and below
by the mass of the resonance. With one or more final state particles
invisible, we shall not be able to find an observable which is bounded
both above and below, but we shall be able to find observables which
are bounded above or below. By finding the observable which gives the
strongest upper or lower bound, we obtain an observable for which the
signal is as peaked as possible, and which may be considered optimal
in the sense of being discoverable in the presence of a smoothly (and
slowly) varying background. What is more, if the location of that peak
(or other sharp feature)
is strongly correlated with some physical property of the signal (such
as the mass of the resonance), then it may be that the same observable
is optimal for the purpose of that measurement.

 We now consider an example in which
this observable is nothing other than the transverse mass.
\subsection{Bounding Variables}	
\subsubsection{$W$ and $W^\prime$ decays}	
Consider decays of a $W$-boson into a charged lepton and an invisible
neutrino.
One way to define the transverse mass, $m_T$,\cite{vanNeerven:1982mz} is as the observable
that provides the greatest lower bound on the ({\em a priori} unknown) mass
of a resonance which decays into some combination of visible and
invisible final state particles. 

The classic example is the decay of a $W$-boson into a charged lepton and an invisible
neutrino, for which the above prescription yields
\begin{gather} \label{mt}
m_T^2 = 2(|p_T| |\slashed{p}_T| - p_T \cdot \slashed{p}_T),
\end{gather}
where $p_T$ is the transverse momentum of the charged lepton and $ \slashed{p}_T$ the
missing transverse momentum.

In this way, $m_T$ is seen to be the optimal variable (in the sense
described above)
to discover the $W$-boson (although it was not employed in that way in the
original analysis\cite{Arnison:1983rp,Banner:1983jy}). In fact, $m_T$
is even better than may appear
on the basis of the above considerations. 
Why? Because $m_T$ is able to separate the signal from some of the backgrounds.
Indeed, there are plenty of ways to arrive at an event
containing a charged lepton, missing transverse momentum and zero or
more jets and each of these is a source of background. For example, every time one
produces a $B$-meson, then ten per cent of the time these decay
leptonically, resulting in precisely the same final state. However,
for these events, $m_T$ cannot exceed the mass of the $B$-meson,
namely a few GeV. Thus, $m_T$ has the additional advantage of
providing a separation between the signal events (which may take
values all the way up to $m_W$) and background events coming from
leptonic $B$-decays, which lie below $m_B$, if well-measured.

If, instead of a charged lepton, the visible particle in the final
state had been a jet, then another dominant source of background would
come from events in which the energy, but not the direction, of the
jet had been mismeasured, leading to apparent, fake, missing
transverse momentum, aligned with the jet. But for such a configuration,
$m_T$ is given, roughly, by the jet mass, which again is typically much smaller
than $m_W$. 

It is important to note that one does not have a theorem to
the effect that there is a clean separation between the signal and all
backgrounds.
In typical final states there will be many background components and not
all of them will enjoy this property. But this does not render the above
observations worthless. The point is that, in any given analysis,
one expects that certain background components will be well
understood while others will be less so. One can also, therefore, try
to proceed by finding an
observable which is insensitive to the background with the largest uncertainty.

One can immediately apply these ideas to searches for new physics at
the LHC, namely to searches for a
$W^\prime$-boson. Now, the dominant background comes from the
$W$-boson itself, but again one is guaranteed a clean separation of
signal and background, at least for well-measured events, since for
the background events $m_T < m_W$. This strategy is exactly that which
is employed in recent LHC searches.\cite{Chatrchyan:2011dx,Aad:2011yg}

As an aside, the fact that $m_T$ gives the greatest lower bound on the
parent particle mass means that it contains all of the information
which follows from the kinematic constraints (namely conservation
of energy-momentum and the mass-shell constraints) alone.\cite{Barr:2009jv}
\subsubsection{Di-leptonic Higgs decays via $WW$ }	
The above approach generalizes straightforwardly to decays of a
resonance into multiple visible particles, with the observable $m_T$
in (\ref{mt}) being replaced by
\begin{gather} \label{mtww}
m_T^2 = m_v^2 + 2(\sqrt{p_T^2 + m_v^2} |\slashed{p}_T| - p_T \cdot \slashed{p}_T),
\end{gather}
where $p_T$ is now the transverse momentum of the visible particle
system and $m_v$ is its measured invariant mass.

For decays with multiple invisible particles, things are a little more
subtle. Na\"{\i}vely, one would define the transverse mass by
\begin{gather} \label{mtww2}
m_T^2 = m_v^2 + m_i^2 + 2(\sqrt{p_T^2 + m_v^2} \sqrt{\slashed{p}_T^2 + m_i^2} - p_T \cdot \slashed{p}_T),
\end{gather}
 but
this is not an observable, since the invariant mass of the invisible
system, $m_i$, is not observable. However, since
$m_T$ in (\ref{mtww2}) is a monotonically increasing function of $m_i$, one immediately
sees that the best lower bound on the mass of the decaying resonance
is given by replacing $m_i$ by its minimum value, namely the sum of
the invisible particle masses, if known, or, if not, by its minimum
possible value, namely zero. 

An immediate application\cite{Barr:2009mx} (see also Refs.~\refcite{Barger:1987re,Glover:1988fn}) is to decays of a heavy (above 135 GeV or so)
SM Higgs boson into two charged leptons and two neutrini via (possibly
virtual) $W$-bosons. The resulting observable, called \mttrue in
\refcite{Barr:2009mx} is just given by the right hand side of (\ref{mtww}) and
is central to state-of-the-art analyses carried out by
ATLAS\cite{ATLAS-CONF-2011-111} and CMS\cite{CMS-PAS-HIG-11-014}. In more detail, ATLAS, for example, performs\cite{ATLAS-CONF-2011-111} a
cut-and-count analysis in the interval $0.75m_h < \mttrue < m_h$, for
varying Higgs mass $m_h$ (after other cuts). In accordance with the
arguments given above, this maximizes the signal contribution in the
cut region compared to the dominant background (which comes from
continuum $WW$ production.)\footnote{Previous analyses used the opening
angle between the two charged leptons, $\Delta \phi_{ll}$, as the figure
of merit. This was based on the observation that, for a spinless
resonance (such as the SM Higgs) of mass $2m_W$ decaying (at
threshold) to $WW$, conservation of angular momentum forces the two
charged leptons to be parallel. Away from threshold, the two charged
leptons remain well-correlated. This is a neat example of an analysis
in which dynamics, rather than kinematics, has been used to motivate
the choice of observable.}

In fact, for Higgs decays in this channel, one can do even better.
For $m_h$ above $2m_W$, both $W$-bosons will typically be produced
on-shell, whilst even for  $m_h$ below $2m_W$, one or other $W$-boson
will typically be produced on-shell, at least for Higgs masses large
enough to give sizable cross-section times branching fraction in the
$WW$ channel.

This observation can be used in the following way. One can define two
new observables, $\mtstar$ and $\mtbound$, as the ones that give the greatest lower bound on the
mass of the (Higgs) signal resonance, subject to the condition that
either one or other (for \mtstar )\cite{Barr:2011ux}, or both (for
\mtbound )\cite{Barr:2011si} of the $W$-bosons be on shell.\footnote{In
  retrospect, history might have been kinder to us if \mttrue,
  \mtstar, and \mtbound had been called $m_T^{W^*W^*}$, $m_T^{W^*W}$, and
  $m_T^{WW}$, respectively.}
In both
cases, since one is applying additional constraints, one obtains a
greater lower bound event-by-event, resulting in an even more sharply
peaked signal distribution. 

Unfortunately, one cannot give an explicit, analytic form for these
two variables, but they can be straighforwardly computed using an
algorithmic computer code. Perhaps unsurprisingly, each
gives a significant improvement only in the region in which the
assumptions that are made in the definition hold to a good
approximation.
\subsection{Higgs decays via $\tau\tau$ }	
The variable $\mtbound$ was first introduced in the context of Higgs
decays to $\tau \tau$,\cite{Barr:2011he} which each, in turn, decay either leptonically, producing two neutrinos, or hadronically, producing only one
neutrino. One apparent disadvantage of this variable for the $\tau
\tau$ channel is that the measured momenta must satisfy the condition
$m_{T2} < m_\tau$ in order that $\mtbound$ be well-defined. This
is guaranteed for well-measured signal events, but in the presence of
detector mismeasurements, events may easily fail to satisfy this
condition. Indeed, na\"{\i}vely, a mismeasurement of $\slashed{p}_T$
by $\delta \slashed{p}_T$ would lead to a similar mismeasurement in
$m_{T2}$; since $\delta \slashed{p}_T$ typically far exceeds $ m_\tau$
at the LHC, one would expect the condition would rarely be satisfied. In
reality, the situation is not so bad,\cite{Barr:2011si} but still one finds that
$\mtbound$ does not exist for about a third of events in simulations,
leading to a significant reduction in statistics.\footnote{One can, of
course, resort to using a different observable, such as (\ref{mtww}) for these
events, so they need not be discarded.}

There are other approximations which might be made for
$\tau \tau$ decays, resulting in other observables. For now, we
restrict our attention to those which are solely concerned with
kinematics, deferring consideration of dynamics to the next section.

Firstly, given
existing limits from LEP on the mass of the Higgs boson, it is likely
that the produced $\tau$s will be significantly boosted in the lab frame,
in which case one could assume that their decay products are collinear.\cite{Ellis:1987xu} 

If one makes this assumption, then in fact there are as many
constraints as unknowns, such that the momenta (and the invariant mass
of the $\tau \tau$ system) can be reconstructed unambiguously. To wit,
the two unknown proportionality constants between  the
visible and invisible momenta in each  $\tau$ decay can both be
determined from the two missing tranverse momentum constraints.
So one ends up not with a
lower bound on the invariant mass of the $\tau \tau$ system, but
its precise value, assuming that things are well-measured.

In practice, this collinear approximation seems to be not such a good
one: the presence of finite detector resolution and acceptance results
in solutions with unphysical energies in a significant number of
events (as many as half in simulations carried out by the experimental
collaborations).

An alternative kinematic strategy for hadronic $\tau$ decays is to use the extra information which
is available in such events from of the location of the
secondary ($\tau$ decay) vertices.\cite{Gripaios:2011jm}  The best measured attributes of
these are their impact parameters, namely the displacements of the
secondary vertices measured in a direction perpendicular to the
visible decay momenta. The invisible momentum in each $\tau$ decay must lie
in the plane of the impact parameter and the visible momentum, leading
to a quartic equation (and a fourfold ambiguity) in the
reconstructed Higgs mass. The problem arises of how to deal with the
resulting discrete ambiguity (which in other cases may be compounded by a combinatorial
ambiguity as to which of multiple, indistinguishable final state particles should be assigned
to which decay or part of a decay. A similar situation (but for cascade decays of
supersymmetric particles) was studied in Ref.~\refcite{Cheng:2008mg}. There it was empirically observed
that the wrong solutions appeared to be
correlated with the right solution. An explanation was given in Ref.~\refcite{Gripaios:2011jm}. In
the specific case of $\tau$ decays considered above, for example, it
was pointed out\cite{Gripaios:2011jm} that, in the limit that the
Higgs mass becomes large compared to the mass of the $\tau$, the quartic equation
reduces to a linear equation, with the four solutions coalesced in a single solution. In the real
world, we are slightly away from the limit, but the solutions remain
close together. The fact that the wrong solutions are correlated with
the right one means that one can simply retain all solutions (indeed
in Ref.~\refcite{Gripaios:2011jm}, it was pointed out that it is even useful to retain complex
solutions, to increase the available statistics): the correlation
means that the ``signal'' (in the form of the correct solution) will
not be overwhelmed, but rather will be reinforced, by the ``background'' coming from the wrong
solutions.
\subsection{Supersymmetry and Dark Matter}	
These ideas can all be extended straightforwardly to events with new
invisible particles, such as the LSP of a supersymmetric theory or
another Dark Matter candidate, rather than common or garden neutrinos. If one is
only interested in kinematics, the only relevant property of an
invisible particle is its mass. As we shall see below, this poses an
important complication when it comes to making mass or other
measurements, since it introduces another unknown to be measured. But
for discovery purposes, the only effect of a massive invisible is that
the scale of the visible particles in the final state is set, roughly
speaking, by the mass difference between the parent particle (or particles) produced
in the hard process and the invisible daughter particles. This is
significant for search strategies of the type discussed above, which
aim for a kinematic separation between signal and background, since
the smaller this mass difference becomes, the more the signal overlaps
with the SM backgrounds. So even heavy (but approximately degenerate)
new particles may be difficult to discover in this way. 
To make progress in such a scenario, one would need to use information
about the dynamics as well. We postpone discussion of this to \S \ref{sec:dynamics}.

The second important complication of supersymmetric theories is that
the parent particles are pair produced, with each decaying to one (or
more) invisible daughter particles. Thus, rather than assume, as we
have done above, that the observed missing transverse momentum can be attributed to
a single decay, we should only assume that it is partitioned in some
way between the two decays.

Once we do so, we can again blindly follow our procedure of asking
which observable gives the greatest event-by-event lower bound on the
mass of the decaying parents. The resulting variable, whose definition
remains algorithmic except in special cases,\cite{Lester:1999tx,Cho:2007dh,Cho:2009ve,Lester:2011nj} goes by the name of
$m_{T2}$.\cite{Lester:1999tx,Barr:2003rg} Just as for $m_T$, $m_{T2}$ gives us a variable which is as
``peaked'' as possible for the signal, leading us to hope that a
significant signal excess can
be identified on top of backgrounds. Even better,
just like $m_T$ it has properties which serendipitously ensure a good
separation between the signal and several of the dominant SM
backgrounds, provided that the mass difference between the parents and
invisible daughters is large.\cite{Barr:2009wu} 
This makes our understanding of the backgrounds that appear in the
signal region and the
systematic uncertainties thereon more robust, enabling a discovery or
exclusion to be made even in the absence of a sharp feature in the
signal distribution.\footnote{Unfortunately, one of the sad facts of
  life is that the more invisible particles one has in an event, the
  less sharp is the signal, essentially because the available energy
  in a collision can be shared out among the particles in many more ways.}
Both general-purpose collaborations have now
presented SUSY searches employing $m_{T2}$.\cite{daCosta:2011qk,CMSmt2}

Again, it follows that $m_{T2}$ captures all of the information that
is contained in the energy-momentum conservation and
mass-shell constraints
alone.\cite{Serna:2008zk,Cheng:2008hk,Barr:2009jv}
This will be important in our discussion of mass measurement, in that
if invisible particle masses cannot be measured using $m_{T2}$, then
they cannot be measured at all using kinematics alone. Happily, they can.

Once particles are pair produced, we must consider the ambiguity that
arises in deciding which of the visible particles in the final state
should be grouped together and associated to the decay of the parent.
This problem is, of course, most acute when the decays are identical. Again,
one can take this ambiguity into account in designing the discovery
observable, by asking which observable gives the greatest lower bound on the
parent mass, after minimizing in addition with respect to the possible
combinations. The resulting variable is called
$m_{TGen}$.\cite{Lester:2007fq} 
A different procedure has been suggested in Ref.~\refcite{Choi:2011ys}.

Finally, one can use a similar trick to deal with ambiguities that
arise in deciding which jets are likely to have come from the final state and
which from radiation in the initial state (insofar as these are
well-defined notions).\cite{Alwall:2009zu}
\subsection{Other applications}	
This simple but far-reaching idea, of optimizing discovery potential
by finding the observable that provides the strongest bound on the mass
of a new particle has also been applied in the context of searches for
charged Higgs bosons in top quark decays\cite{Gross:2009wia} and for pair-produced third-generation
leptoquarks\cite{Gripaios:2009dq} decaying to a $b \nu t\tau$ final state.\cite{Gripaios:2010hv}
\subsection{Singularity Variables}	
The kinematic ideas discussed above could, in principle, be applied to
events with arbitrarily complicated topologies. We have already seen, however,
that even in relatively
simple cases, such as $h \rightarrow WW \rightarrow 2l 2\nu$, it is
not possible to find an explicit, analytic expression for the
observable that encodes all of the information in kinematics. 

Kim\cite{Kim:2009si} has noted that the notion of a bounding variable
arises because of endpoint singularities that appear when one projects the full
phase space of some event topology onto its observable subspace (a
simple example was given in Ref.~\refcite{Brooijmans:2010tn}, in which the full phase space is a sphere
and the observable phase space is a disk, singular on its edge).
He has noted, furthermore, that these singularities show more general
behaviour than simply endpoints (such as
cusps\cite{Han:2009ss,Agashe:2010gt}); these more general
features may also provide a way to identify a signal on top of a
smoothly-varying background. Once the singularities corresponding to a
given signal topology have been found, one may try to construct an
optimal observable in the region of the singularity, according to
various criteria of optimality. De R\'{u}jula
and Galindo\cite{Rujula:2011qn} have attacked this problem in the
simplest case of $W$-boson decays, defining observables
which are optimal in a statistical sense.
\section{Dynamics \label{sec:dynamics}}	
All of the methods dicussed above use only the kinematic properties of
the signal and backgrounds. We now discuss analyses which try, either
separately or in addition, to exploit what is known, or may be
assumed, about the dynamics.  
\subsection{Higgs decays via $\tau\tau$, again }	
For $h \rightarrow \tau \tau$ searches (either in the SM or MSSM),
both ATLAS\cite{ATLAS-CONF-2011-132} and CMS\cite{Chatrchyan:2011nx} have employed more sophisticated approaches using 
{\em ad hoc} likelihood functions. Neither of the approaches
implemented so far uses information
from the secondary vertices, such that the kinematics of the signal
are underconstrained. The analyses are similar in that they both allow
the unconstrained parameters, namely the momenta and invariant masses
(for hadronic decays) of the invisible systems in each $\tau$ decay to
vary, subject to the mass shell and missing transverse momentum
constraints. The Higgs mass is then reconstructed by maximizing a
likelihood function on this space, whose nature differs in the two
analyses. The ATLAS likelihood function\cite{Elagin:2010aw} contains two ingredients. The
first is a probability density function for the $\tau$ decay angles,
which is obtained by fitting a simulation of $Z \rightarrow \tau \tau$
decays. The second ingredient is a Gaussian smearing of the measured
missing transverse momentum.\cite{ATLAS-CONF-2011-080}
CMS are more reticent about the details of their likelihood function;
we are at least told that it
``[takes] into account the tau-decay phase space and the probability density in the tau transverse momentum, parametrized as a function of the tau-pair mass''.\cite{Chatrchyan:2011nx}
\subsection{Supersymmetry and Dark Matter, again}	
There have been several, rather different, efforts to use dynamical
information to optimize discovery of theories with pair production,
such as
supersymmetry. The most basic of these continue the spirit of the
kinematic discussion above, but either supplement or substitute it
with dynamical assumptions. 

As an example, one might make the dynamical assumption that, because
parton distribution functions fall off steeply with increasing $x$,
parents will tend to be pair produced at rest relative to one another
in their CM frame, so as to minimize the resulting $\hat{s}$. If one makes this
assumption, then the natural observable is defined in
Ref.~\refcite{Rogan:2010kb}, where it is denoted $M_R$. Since pairs of
squarks are produced in a $p$-wave state at leading order in QCD in quark-antiquark collisions, it
is clear that this assumption cannot be valid in certain cases. But,
as discussed above, this does not render the observable invalid;
rather, it only makes it possibly
sub-optimal.\footnote{Ref. ~\refcite{Rogan:2010kb} also introduced a
 modified variable $M_{R^*}$, designed to deal with this issue.}

It certainly is true, however, that pairs of parents are produced back-to-back
in their CM frame. If one made, furthermore, made the dynamical
assumption that the CM frame of the parents was dominantly boosted only
longitudinally with respect to the colliding beams, then the optimal
observable might be one which was invariant under both longitudinal
boosts of the CM frame of the parents and back-to-back boosts of the
parents in their CM frame. Indeed, one would expect the bulk of
signal events to occur at the same value of such an observable (in
exactly the same way that Lorentz invariance guarantees a peak in the
invariant mass in decays without missing transverse momentum). The
appropriate variable goes by the name of $m_{CT}$.\cite{Tovey:2008ui}

These approaches could be applied to arbitrary pair decay
processes. For longer cascade decay chains, we have seen that it is
kinematically possible to reconstruct events. However, one might
wonder how reliable such a reconstruction is in the presence of
detector mismeasurements. One way to deal with this is to construct a likelihood which
takes the detector response into account.\cite{Allanach:2004ub,Webber:2009vm,Nojiri:2010dk}

 A completely different approach deals with the aforementioned
difficulty of discovering SUSY using
naive kinematic observables if the superpartner spectrum is roughly
degenerate, or ``compressed''.

At least if the superpartners are heavy, one important feature of the signal is that it leads to a harder pattern
of radiation in the initial state than what is typically obtained for
SM backgrounds. Thus, the optimal way to discover such scenarios may
be to focus on events with more jets than expected from the signal
final state and with large
missing transverse momentum, arising when the parents recoil strongly against the
initial state radiation.\cite{Alwall:2008ve,Alwall:2008va} By incorporating a number of selection cuts,
one can achieve a good coverage of the signal parameter space, without
the need for sophisticated observables.\cite{Alves:2011sq} One can
also try to tag initial state radiation explicitly.\cite{Krohn:2011zp}

Yet another direction is to focus instead on the dynamics of the
background rather than that of the signal. So far, efforts have been
directed at the dominant background of mismeasured QCD events.
Here, the presence of the background is purely a detector effect of
which one can hope to build up a good understanding. As a simple
example, the fact that the largest mismeasurements come from jet
energies (rather than jet directions) means that one can suppress much
of this background by cutting on an observable with known properties
under such mismeasurements. One such observable, $\alpha_T$,\cite{Randall:2008rw} works well for
di-jet events, though the
signal rejection is high. More challenging are events with multiple
jets, since our understanding of the SM background prediction is very
poor. ATLAS\cite{ATLASmulti} has performed a search in events with between six and
eight jets in which the SM background is determined by extrapolation
from measurements in events with fewer jets. The key observable is
$|\slashed{p}_T|/\sqrt{H_T}$,
where $H_T$ is the scalar sum of the transverse momenta of all jets
within some fiducial region. This ratio measures the missing
transverse momentum relative to the resolution due to stochastic
variations in the measured jet energies that are assumed to follow the
central limit theorem. It is then expected that the distribution of
this observable will be roughly invariant under changes in the jet
multiplicity, such that the background in the event sample with high
jet multiplicity can be estimated from the measured distribution in
the low jet multiplicity sample, which is assumed to be background
dominated.
A similar search was carried out by CDF,\cite{Aaltonen:2011na} using a
cut on this variable. CDF's search also employs a technique in which the
total missing transverse momentum is compared with that observed using
the charged particle spectrometer only, the rationale being that these
should be aligned for signal events, but may be anti-aligned for
background events arising from calorimeter jet energy mismeasurements.

Finally, one could attempt a more rigorous likelihood method using
genuine matrix elements for signal process. Though this has not been
explored in the context of discovery, a study of the prospects for mass
measurement has been carried out.\cite{Alwall:2009sv,Alwall:2010cq}
\section{Measuring New Physics\label{sec:measurements}}	
We now turn to a discussion of methods that have been developed to
measure the properties of new physics, once it has been
discovered. As we remarked in the introduction, it is clear that the
process of making measurements is not wholly divorced from the
process of making a discovery, just as the process of excluding
models currently underway at the LHC depends on where in the parameter space one sits.
Indeed, many of the observables and stategies proposed for making
measurements coincide (or nearly coincide) with discovery
strategies. This is hardly surprising, in that observables which do a
good job of distinguishing signals from backgrounds will also
facilitate making clean measurements.

Furthermore, the topics of mass and spin measurements have already
been reviewed in the recent literature.\cite{Wang:2008sw,DeRoeck:2009id,Barr:2010zj} Our discussion will,
therefore, be brief.
\subsection{Mass measurements \label{sec:masses}}
The first measurement priority will be to establish the kinematic
properties of the signal, namely the masses of any new particles.
Just like the discovery process, this is complicated by the fact that
kinematic information is lost in events with invisible particles. 

It has long been known (beginning with the pioneering study of
endpoints in invariant mass distributions in Ref.~\refcite{Paige:1996nx}) that
some information about particle masses (roughly speaking, mass
differences) can
be obtained using kinematics alone. 

In events involving cascade decays, with on-shell intermediate states,
it may be possible, as discussed above, to reconstruct events. This
approach was pioneered in Ref.~\refcite{Nojiri:2003tu} and has been developed
in
Refs.~\refcite{Kawagoe:2004rz,Allanach:2004ub,Cheng:2008mg,Cheng:2009fw,Webber:2009vm,Nojiri:2010dk,Gripaios:2011jm}.
The business of counting the various unknowns and constraints is
straightforward and we shall not describe it here. As an example of
how many constraints is required, for pair decays one needs two, on-shell
intermediates in each chain in order to be able to reconstruct the masses.

A crucial question is whether masses can be measured absolutely in decays with
fewer on-shell intermediates, using kinematics alone. A definitive
answer to this has been provided as follows: when a parent particle
(or a pair of identical parent particles) decays into a system of visible
daughter particles and a system of invisible daughter particles, there is enough
information in kinematics to measure the mass of the parent and the
sum of the masses of the invisible daughters, even when there are no
on-shell intermediates.\cite{Cho:2007qv,Gripaios:2007is,Barr:2007hy,Cho:2007dh}
In particular, if a decay process involves only a single invisible
dark matter particle, then the mass of that particle, can, in
principle, be measured, no matter what decay produces that particle.

What has not been answered definitively is the question of how best to
measure those masses, or indeed whether it can be achieved in
practice. Of course, the answer to this question depends to a large
extent to the nature of the signal and many different methods have been
put forward. These have been painstakingly reviewed in Ref.~\refcite{Barr:2010zj}, to
which the reader is referred for details. 

Regrettably, along with this plethora of methods has come a {\em sm\"{o}rg\r{a}sbord}
of different observables with confusingly similar
names. Ref.~\refcite{Barr:2011xt}
counted over fifty different variables in the literature which go
under the name of ``transverse mass'', with or without further
qualification.
This state of affairs is probably unavoidable; in any case, there
would seem to be an onus
upon authors and workers in the field to take care over the definition of observables.
\subsection{Spin measurements \label{sec:spins}}
Spin measurements are challenging even in the absence of missing
transverse momentum, because spins (being the generators of rotations) 
manifest themselves in angular effects. In order for such effects to
be present in a decay, for example, one first needs to create a
preferred axis of some kind. So to have observable spin effects often
requires a complicated event topology. For example, in a cascade decay, in order that non-trivial angular correlations
exist, one requires at least that the intermediate particle in
question be polarized, and if a fermion, that its decay be chiral.\cite{Wang:2006hk}

Things are even more difficult given the presence of missing
transverse momentum, because one needs to be able to reconstruct
reference frames in order to be able to measure angles
explicitly. One way to circumvent this problem is to focus on
Lorentz-invariant quantities, which take the same value in all
frames. Alternatively, one needs to reconstruct events, which can only
be achieved for certain topologies, as described above.

There are, nonetheless, many ideas on the market, most of which apply
only to a specific decay topology. They are reviewed in
Refs.~\refcite{Wang:2008sw,DeRoeck:2009id}.
\section{Other measurements \label{sec:other}}
Besides measuring masses and spins, one would eventually like to
measure all of the parameters appearing in the lagrangian that
describes physics at the TeV scale. These include the gauge and global
charges of
new particles (with respect to whatever symmetry group is present at
that scale),
as well as the coupling strengths of the interactions that are
permitted by that symmetry group. The method by which one might
achieve this is, of course, greatly contingent upon the nature of the
new physics and relatively little work has been appeared in the
literature. 

To give just one example of what has been done, Ref.~\refcite{Burns:2008cp} explored in detail the procedure by which
one might measure couplings and mixing angles, in addition to masses
and spins, in cascade decays using
invariant mass observables.

We now describe a rather different measurement that has been proposed
recently,\cite{Giudice:2011ib} namely to count invisible particles in events.
\subsection{Counting invisible particles \label{sec:cunt}}
If we are fortunate enough to discover new physics associated with
missing transverse momentum at the LHC, there will surely be a great temptation to
identify it with the Dark Matter that dominates the matter in the
cosmos.

The question arises, though, of how to make that identification
concrete. The holy grail would, presumably, be to compute the expected
relic density using fundamental parameters of the Lagrangian measured
at the LHC or in other terrestrial experiments and to compare with the
observed cosmic abundance.

Unfortunately, it would appear that such a comparison is out of reach
in the LHC era.\cite{Baltz:2006fm} For one thing, such a comparison would require one
to measure a large number of parameters, each one of which would, from
the previous discussion, require a herculean effort at the
LHC. Indeed, a computation of the relic abundance would require one
not only to discover all the new particles appearing at the TeV scale,
but also to determine their masses, spins, and couplings to all other
particles. For another, the observed relic abundance implies that it
is the weak interaction that is responsible for setting the dark
matter density, but weak interaction effects are inevitably
overwhelmed by strong interaction effects at a hadron collider and so
near impossible to disentangle. As an explicit example, cross sections
for processes involving production of new particles followed by their
decay are set in the narrow width approximation by the product of the
production cross section and the branching fraction for the decay, but
the production cross-section will usually be dominated by the strong
interaction.

Ref.~\refcite{Giudice:2011ib} made the rather different suggestion of
trying to gain information about a Dark Matter candidate by simply
counting invisible particles in events with missing energy. Indeed,
multiple production would already constitute evidence for the presence
of a symmetry, as required to stabilize Dark Matter on Hubble
timescales.
Similarly, if one could establish evidence for production of an odd
number of invisible particles, one could rule out the simplest (and
most commonly employed in models) symmetry, namely $Z_2$. 

But how can invisible particles be counted in LHC events? The
available energy in collisions is, of course, shared out in a random
fashion between the particles produced and, as a result, the shapes of
various observable distributions vary depending on how many invisible
particles are present. As a simple example, if a parent particle
is singly produced at the LHC and decays to visible particles and
$n$-invisible particles whose masses may be neglected on collider
scales, then the transverse mass (\ref{mt}) has a power-law behaviour
near its maximum, with the power given by
$n-\frac{3}{2}$.\cite{Giudice:2011ib} This result holds independently
of the number of visible particles, or indeed of whether there are
intermediate particles on-shell. At least in this simple example, it should be
rather easy to distinguish between production of a single, invisible
particle (for which the distribution diverges near the endpoint) from
production of multiple, invisible particles (for which the
distribution vanishes). 
\subsection{Distinguishing stabilization symmetries}	
A related question is how we might try to distinguish between different
stabilizing symmetries for Dark Matter. Consider, for example, the two
simplest symmetry groups, namely $Z_2$ and $Z_3$. Counting invisible
particles provides one way, in that an odd number of invisible
particles can only be associated with $Z_3$ and not $Z_2$. Previous
work\cite{Walker:2009ei,Agashe:2010gt,Agashe:2010tu} attacked this question in a different way, focussing on the
observation that models with $Z_3$ symmetries might contain final
states with the same visible particle content, but with different
numbers of invisible particles, which would lead to distributions
containing two, rather than one, components, which might be picked out in
data.
\section{Characterization of new physics \label{sec:simplified}}	
Even once we have made a discovery of new physics and possibly
measured a limited number of its properties, we shall be faced the question of how
best to try to characterize that physics. The holy grail is, of
course, to write down the Lagrangian, in which the fundamental
principles of symmetry, etc., are as manifest as we know how to make
them.

Unfortunately, it seems unlikely that we will succeed in
guessing the correct form of the Lagrangian soon after discovering new
physics. Indeed, it may not be possible at all to home in on the
Lagrangian in the LHC era, let alone to perform precise measurements
of its parameters. 

But some characterization of the new physics would be invaluable in
terms of suggesting where to look for corroborations of discoveries,
to perform cross-checks, or indeed
to further enhance our knowledge about the nature of the physics in a
sort of bootstrap process.

There seem to be two possible approaches. The first is make an
educated guess (perhaps on the basis of one's theoretical bias) as to what the exact Lagrangian is, modulo the values
of a few parameters. One can then either exclude the model in an
increasingly large region of parameter space (in the absence of a
signal excess) or attempt to measure those parameters (in the presence of
one).

This approach has been very popular, understandably so since it gives
one a concrete hypothesis which (one hopes) will eventually be
accepted or rejected. One downside is that one's guess must be a
good one and, though it is a tale which is too long for us to tell
here, our theoretical bias has not in the recent past proven to be a very good indicator
for what new physics does lie beyond the Standard Model. A perhaps more
pernicious downside is that focussing on a specific Lagrangian can
lead to the development of search strategies which, though optimized
for that particular Lagrangian, fail spectacularly when one considers
other new physics, that is not dramatically different from the
original Lagrangian. A salutary example is provided by supersymmetric models with
unification of gaugino masses at the GUT scale. The RG evolution leads
to a fixed, large (6:1) mass ratio between the gluino and the Bino at the TeV
scale and, since the Bino is often the LSP in such models, a fixed
signal kinematics. Searches which were developed with such models in
mind have proved to be rather ineffective in probing models with
compressed spectra.\cite{Alwall:2008ve,Alwall:2008va}

The second approach avoids this by via a more {\em ad hoc}
strategy of making a coarse guess for new physics in the form of a
simplified model\cite{ArkaniHamed:2007fw,Alves:2011wf} which, it is hoped, captures the essence of the
physics that is relevant for LHC collisions. This can hardly be described
as a novel philosophy, in that physics has been done in much this way for
millennia, but the details might be. Typically, a simplified model might contain only a few new
degrees of freedom, with masses allowed to float as well as the
cross-sections for production and decay. The advantage of such an
approach is that it enables a huge variety of models to be rapidly
scanned and compared with LHC data; the disadvantage is that such a model is in no
way fundamental and it is hard to see how it could be used
to make detailed predictions for new physics phenomena elsewhere, for
example in the cosmos. Presumably though, once one had narrowed down the
possibilities to a small set of simplified models, one could return to
the approach of guessing concrete Lagrangians, for which this can be done.
\section*{Acknowledgments}
I am greatly indebted to A.~J.~Barr and C.~G.~Lester, who were
instrumental in inculcating in me much of the material discussed. I
also thank M.~Pierini for correspondence regarding $M_{R^*}$.
\providecommand{\href}[2]{#2}\begingroup\raggedright\endgroup
\end{document}